\documentclass[a4paper,fleqn]{cas-sc}

\usepackage{float}
\usepackage{placeins}
\usepackage{graphicx}
\usepackage{lineno}
\usepackage[numbers]{natbib}
\usepackage{caption}
\usepackage{subcaption}
\usepackage{float}

\usepackage[normalem]{ulem}
\usepackage{xcolor}

\def\tsc#1{\csdef{#1}{\textsc{\lowercase{#1}}\xspace}}
\tsc{WGM}
\tsc{QE}

\begin{document}

\let\WriteBookmarks\relax
\def\floatpagepagefraction{1}
\def\textpagefraction{.001}

\shorttitle{A cosmic muon trigger for the sMDT detector}    

\shortauthors{Li et al.}  

\title [mode = title]{Fast Self-Triggering and Segment-Fitting of Cosmic Muons with a small-Diameter Monitored Drift Tube Detector}  



%


\author[1]{Shuzhe Li}[orcid=0009-0000-7953-5590]
\ead{shuzheli@umich.edu}

\credit{<Credit authorship details>}

\affiliation[1]{organization={Department of Physics, University of Michigan},
	city={Ann Arbor},
	state={MI},
	postcode={48109},
	country={USA}}
\cormark[1]	

\author[1]{Yuxiang Guo}[orcid=0000-0002-6027-5132]
\ead{gyuxiang@umich.edu}

\author[1]{Jiajin Ge}[orcid=0000-0002-9204-4718]
\ead{jiajinge@umich.edu}

\author[1]{Chihao Li}[orcid=0000-0002-7931-6504]
\ead{chihao@umich.edu}

\author[1]{Bing Zhou}[orcid=0000-0002-0034-6576]
\ead{bzhou@umich.edu}

\author[1]{Junjie Zhu}[orcid=0000-0002-5278-2855]
\ead{junjie@umich.edu}


\cortext[1]{Corresponding authors.}


\begin{abstract}
	We present a fast self-triggering and segment-fitting algorithm that identifies cosmic muons and reconstructs their directions using a small-diameter Monitored Drift Tube (sMDT) chamber without an external trigger detector. The muon arrival time is determined using a Legendre-based algorithm combined with coarse- and fine-grid searches. We implement the algorithm in FPGA firmware on an ATLAS MDT miniDAQ board and evaluate its efficiency, resolution, and latency. The results demonstrate standalone sMDT triggering and trigger-level reconstruction of muon directions, providing a solution for chamber commissioning and potentially MDT-based triggers for charged massive stable particle searches at the High Luminosity-LHC.
\end{abstract}


\maketitle

\section{Introduction}\label{sec-intro}
Drift-tube detectors are widely used for precision tracking in particle and nuclear physics experiments, with several recent applications documented in~\cite{BaBar:2001yhh, Hirano:2000ei, Belle-II:2010dht, VanHaarlem:2010yq, Danielsson:2013qta, Mu2e:2014fns, Johansson:2023voo, CMS:2009szg, Bauer:2001scz, DUNE:2021tad}. When a charged particle traverses a drift tube at time $t_0$, electrons produced in primary and secondary ionizations drift toward the central wire. The time difference between the earliest signal arrival at the wire ($t_s$) and the reference time ($t_0$) defines the hit drift time, $t_{\mathrm{drift}} = t_s - t_0$, which is then converted into a drift radius using a time-to-radius ($R(t)$) relation\cite{Nelson_2021}. In a multilayer chamber, the hit tubes provide multiple drift circle measurements, and the particle trajectory is reconstructed as the line tangent to these circles. Drift tube detectors typically achieve a spatial resolution of approximately 100 $\mu$m per tube, with drift times ranging from 100 ns to 1 $\mu$s, depending on the gas mixture, operating pressure, and applied high voltage.

In the NA62 experiment, the straw tube tracker was integrated into the first-level trigger system by exploiting the similar arrival times of electrons produced near the walls of tubes traversed by the same charged particle~\cite{Danielsson:2013qta}. However, because drift tube detectors have relatively long drift times and cannot readily determine $t_0$ independently, they are generally not used for triggering in collider experiments and specific trigger detectors are needed. For example, for the planned ATLAS upgrade for the High-Luminosity LHC (HL-LHC)~\cite{Gkountoumis:2026ekw}, the barrel resistive plate chamber (RPC) and endcap thin-gap chamber (TGC) trigger detectors will provide $t_0$ measurements and define regions of interest for identifying tubes that may have been traversed by charged particles.  A fast segment-finding and track-fitting algorithm will then process the selected tubes to estimate the charged particle’s momentum at the first-level trigger. This approach reduces the number of candidate hits that must be analyzed in real time and allows the trigger system to incorporate precise tracking information from the drift tube chambers within its stringent latency constraints.

This article presents the development of a fast self-triggering and segment-fitting algorithm to trigger on cosmic muon events and determine their directions in real time using a Legendre-transform-based method. The algorithm is implemented in firmware and executed on the field-programmable gate array (FPGA) of a mini data-acquisition (miniDAQ) system~\cite{Guo:2022xye} to collect cosmic muon events with an ATLAS small-diameter Monitored Drift Tube (sMDT) detector~\cite{Wei:2022gmi, Amidei:2022jlx}. The algorithm's performance is evaluated in terms of trigger efficiency, segment-direction and $t_0$ resolutions, spatial resolution per tube, and latency. 

Although the cosmic muon event rate is substantially lower than the event rate encountered under LHC collision conditions, self-triggering on cosmic muons presents significant challenges. Cosmic muons can traverse the chamber at arbitrary angles, whereas muons produced in LHC collisions typically originate from the interaction point. In addition, cosmic muons arrive asynchronously and may traverse the detector at any time relative to the system clock. By contrast, muons at the LHC are produced in $pp$ collisions synchronized with the accelerator bunch crossings, and their presence is identified by dedicated trigger devices. 
For cosmic muon events, no external timing reference is available, and $t_0$ must instead be inferred directly from the drift-tube signals in real time. Our algorithm addresses this challenge by simultaneously identifying a consistent pattern of tube hits, estimating the muon arrival time, and reconstructing the corresponding track segment. This capability provides a low-latency solution for standalone cosmic muon triggering and may facilitate detector commissioning, calibration, testing, and performance studies when external trigger devices, such as dedicated trigger chambers or scintillator detectors, are unavailable. It also has the potential to be used to trigger on slow-moving charged particles produced in collider experiments. 

\section{Experimental setup}
An ATLAS sMDT prototype detector is used to collect cosmic muon events and evaluate the performance of the self-triggering algorithm. The detector comprised four sMDT chambers (only two are used in this study), each containing 48 cylindrical aluminum drift tubes arranged in four layers.  Each tube had a diameter of 1.5 cm and a length of 40 cm. The four chambers are positioned in the $x$-$y$ plane, with adjacent chambers oriented orthogonally to each other. The drift tubes were operated with an Ar:CO$_2$ (97:3) gas mixture at an absolute pressure of 3 bar. A high voltage of 2,730 V was applied. Under these operating conditions, the electron maximum drift time is around 190 ns and the single-tube spatial resolution is approximately 100~$\mu$m. The noise rate is below 50 Hz per tube with a threshold of 27 mV. 

The sMDT tubes are instrumented with the new ATLAS front-end electronics developed for operation at the HL-LHC. Signals from each tube are amplified, shaped, and discriminated by an Amplifier/Shaper/Discriminator (ASD) ASIC~\cite{DeMatteis:2017xky}. To suppress multiple threshold crossings of the same detector pulse, the ASD imposes a programmable dead time of approximately 900 ns. After the first threshold crossing is registered, subsequent crossings occurring within this interval are ignored. In addition, the ASD integrates the signal charge over approximately the first 20 ns. The pulse amplitude is determined using a Wilkinson-type conversion implemented within the ASD. The charge accumulated on the integrator capacitor is discharged at a constant rate, converting the measured charge into a time interval. The leading and trailing edges of the ASD output pulse are digitized by a time-to-digital converter (TDC) ASIC with a time-bin width of 0.78 ns~\cite{Wang:2017jnd, Liang:2019weg}. Unlike the front-end electronics operated from 2010 to 2026, in which hits were buffered within the TDC and read out only upon receipt of a trigger signal, the new front-end electronics operates in a triggerless mode. It continuously transmits all recorded muon hits off the chambers to the new trigger and readout system for real-time processing. This new mode of operation enables the sMDT detector to self-trigger on cosmic muons without requiring an external trigger device.

Each ASD ASIC process the signals from eight drift tubes, while a single TDC receives and digitizes the discriminator outputs from three ASD ASICs. Each mezzanine card, consisting of three ASDs and one TDC, reads out 24 tubes. A Chamber Service Module (CSM)~\cite{Teng:2026jaf} aggregates and multiplexes data from up to 20 mezzanine cards. The combined data stream is transmitted to a miniDAQ board~\cite{Guo:2022xye}, where the raw data are decoded and the hits are selected. 
The miniDAQ board also configures the front-end electronics and executes the self-triggering algorithm, which triggers on cosmic muons and determines their arrival times and directions in real time.

A scintillator is positioned 33 cm below the sMDT detector and is used in two distinct configurations. In the first configuration, cosmic muon events are triggered by the sMDT self-triggering algorithm, while the scintillator provides an independent time measurement to determine the muon arrival time, without participating in the trigger decision. In the second configuration, a separate data-taking run is performed in which the scintillator generates the event trigger.

\begin{figure}[!htbp]
    \centering
    \subfloat[]{\includegraphics[width=0.35\textwidth]{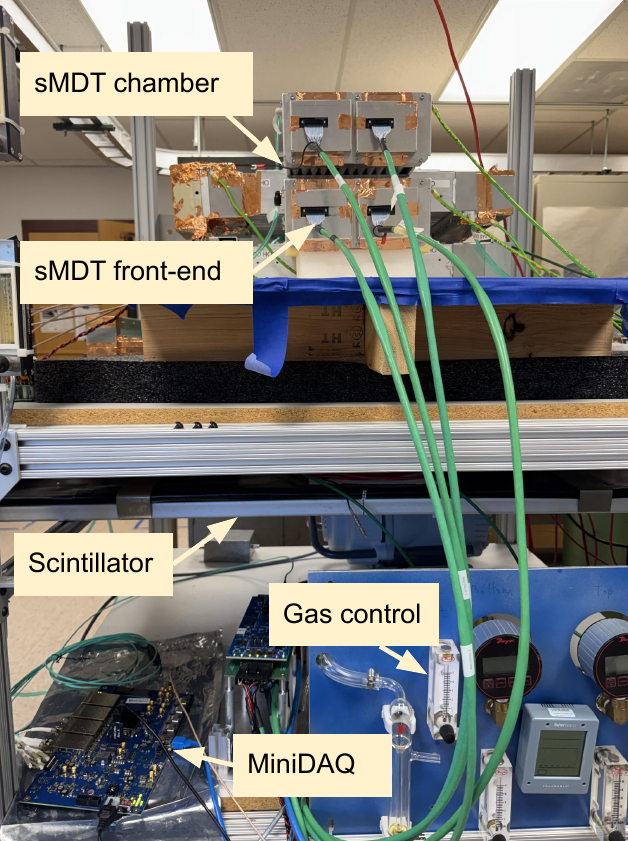} \label{fig:geometry} } 
    \subfloat[]{\includegraphics[width=0.25\textwidth]{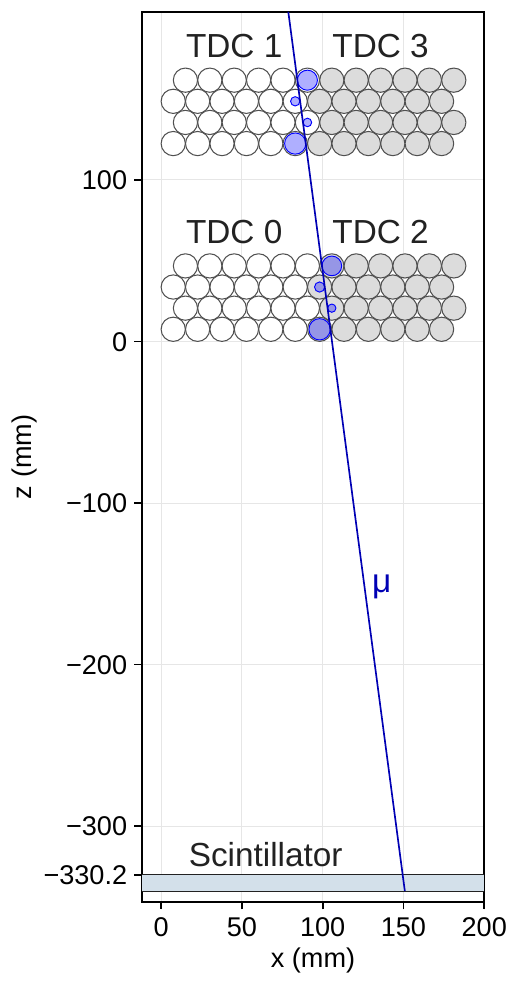} \label{fig:resolution} } 
    \caption{(a) Experimental setup in the lab; and (b) The schematic layout of the setup.}
    \label{fig:setup}
\end{figure}

Figure~\ref{fig:setup} shows the overall experimental setup. 

\section{Self-triggering and segment-fitting algorithms}
\subsection{Legendre-transform track-finding algorithm with $t_0$ vote}

\begin{figure}[ht]
    \centering

    \begin{subfigure}[b]{0.45\textwidth}
        \centering
        \includegraphics[width=\textwidth]{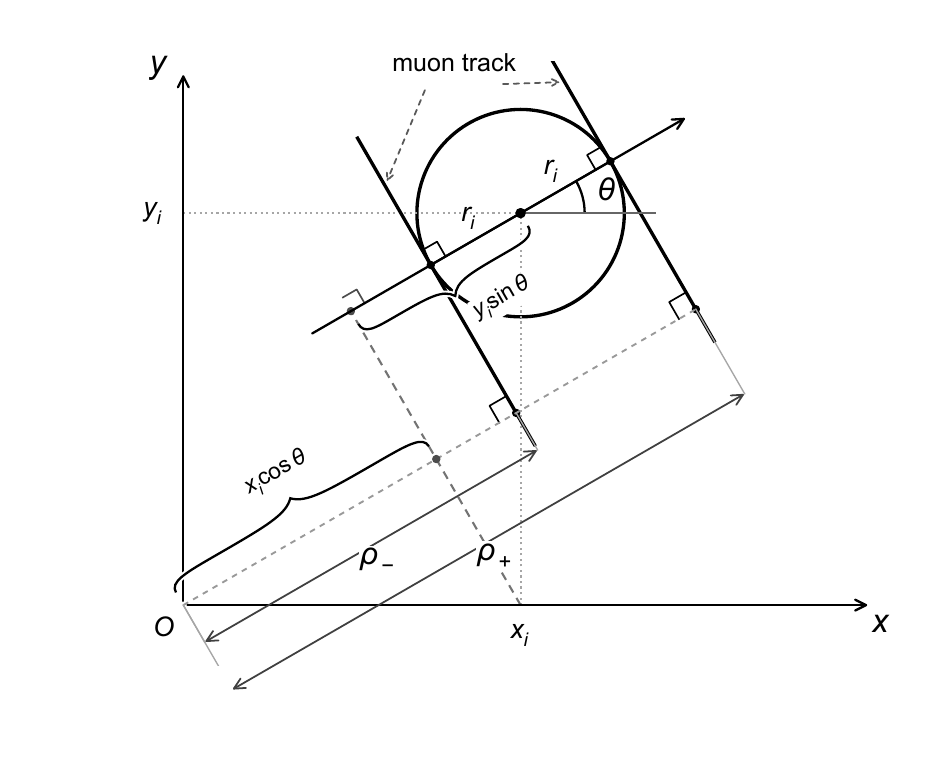}
        \caption{}
        \label{fig:geometry}
    \end{subfigure}
    \begin{subfigure}[b]{0.45\textwidth}
        \centering
        \includegraphics[width=\textwidth]{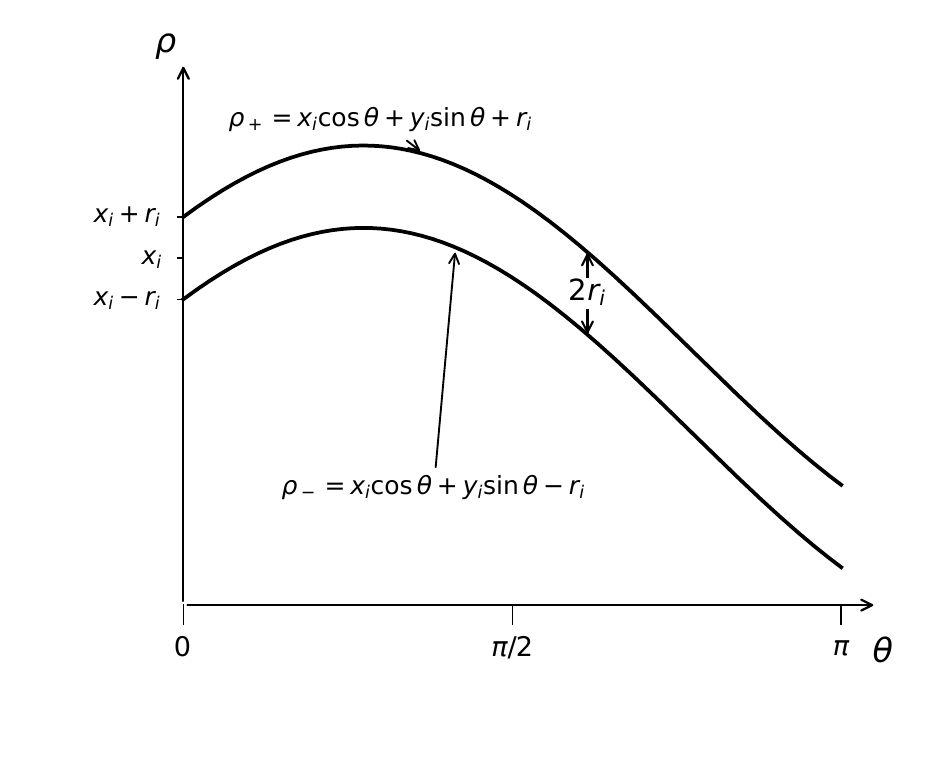}
        \caption{}
        \label{fig:legendre}
    \end{subfigure}

    \caption{Legendre transform of the drift circle for a single tube with the wire at $(x_i, y_i)$: (a) Compatible tracks at angle $\theta$; and (b) All possible tracks in Legendre space.}
    \label{fig:track_reconstruction}
\end{figure}

A Legendre-transform algorithm~\cite{Alexopoulos:2008zza} is used to find all tubes hit by the same muon. The algorithm reconstructs a straight-line segment by mapping each drift tube hit into curves in a parameter space defined by the track parameters $(\theta,\rho)$, where $\theta$ specifies the direction normal to the track and $\rho$ is the track’s perpendicular distance from the origin. As illustrated in Fig.~\ref{fig:geometry}, for a wire located at $(x_i,y_i)$ with measured drift radius $r_i$, the compatible track parameters satisfy $\rho=x_i\cos\theta+y_i\sin\theta\pm r_i$, where the two signs reflect the left–right ambiguity. As $\theta$ is scanned over $[0, \pi)$, each drift circle therefore maps onto two sinusoidal curves in the $(\theta,\rho)$ space, as shown in Fig.~\ref{fig:legendre}. For hits produced by the same muon, these curves converge near a common point, yielding a peak in the discretized parameter-space accumulator. The location of this peak determines the position and direction of the reconstructed muon track segment.

To determine the segment position and direction from the peak in $(\theta,\rho)$ space, the drift radius $r_i$ must be known for each hit tube. Unlike collision events, which have a well-defined time reference determined by the trigger devices, cosmic muon events have an unknown arrival time that must be estimated during segment reconstruction. Accurately estimating $t_0$ is the main challenge of this work. 

Exhaustively scanning $t_0$ and applying the Legendre transform at each step would be computationally expensive. Instead, we evaluate candidate tracks with a $t_0$-voting method. For each candidate track, we estimate each hit's drift radius as the perpendicular distance from its wire center to the candidate track and convert it to a drift time using the inverse time-to-radius relation $R(t)$. Subtracting this drift time from the measured hit time $t_i$ gives a per-hit estimate of $t_0$. The hits then vote for a common $t_0$, and the consistency of their estimates determines the candidate track’s score.

The procedure is as follows. We first apply the Legendre transform to the wire-center positions of the hit tubes to construct a $(\theta,\rho)$ histogram. Its peak provides an initial track estimate. Coarse- and fine-grid searches around this peak then evaluate additional track candidates through voting for a common $t_0$. Once a valid $t_0$ is identified, the drift radius of each hit tube is calculated as $r_i=R(t_i-t_0)$. Using these radii, we repeat the Legendre transform over a smaller region of $(\theta,\rho)$ space with finer binning. The peak of this refined histogram determines the final track parameters.

The Legendre transform is therefore only applied twice: first to the wire-center positions for initialization, and then to the drift radii for final track refinement. The intervening grid searches use inverse-$R(t)$ calculations and $t_0$ voting. These candidate evaluations and voting operations are well suited for parallel implementation in the FPGA firmware.

\subsection{Firmware implementation in FPGA}
\begin{figure}[pos=ht]
     \centering
     \includegraphics[width=0.4\linewidth]{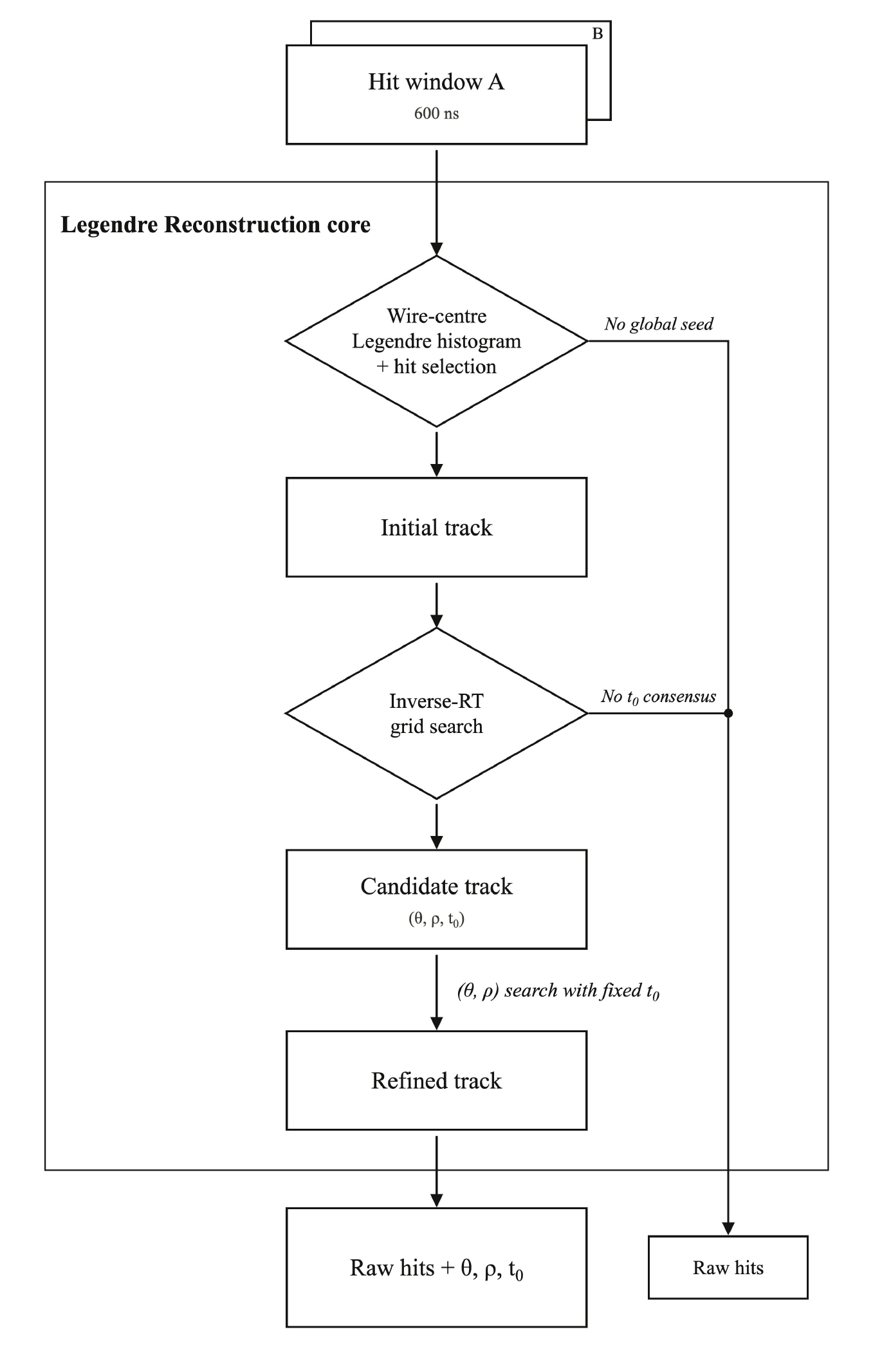}
     \caption{Overall data-processing pipeline.}
     \label{fig:overall_pipeline}
 \end{figure}

The algorithm is implemented on the Xilinx Ultrascale XCKU035 FPGA of the miniDAQ board. Figure~\ref{fig:overall_pipeline} shows the overall schematic of the self-triggering algorithm. Hits received from the front-end electronics are buffered in two FIFO banks that work in ping-pong mode. The first hit in an empty bank initiates a user-configurable acquisition window, and all subsequent hits arriving within this window form an event candidate. Only candidates that meet a user-defined hit-multiplicity threshold are processed for segment-fitting, and candidates below the threshold are discarded. Hit accumulation for reconstruction proceeds concurrently with event acquisition, and segment-fitting begins once the acquisition window closes. 

\begin{figure}[pos=ht]
    \centering
    \includegraphics[width=\linewidth]{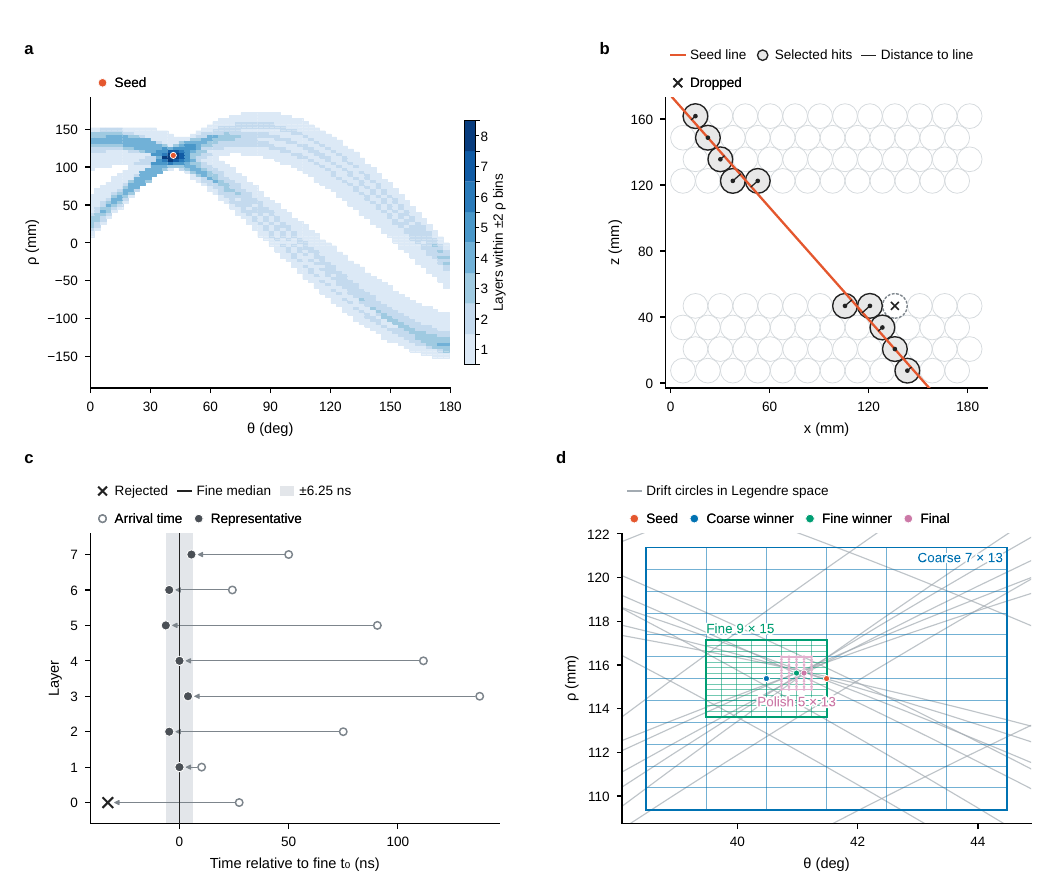}
    \caption{Example of reconstruction steps on a recorded event. (a) Wire-center Legendre histogram. The seed is determined by the centroid of the peaks in the histogram (b) Hits selection. For each layer, only two closest hits are retained. Hit crossed out in the figure is dropped. (c) $t_0$ consensus by inverse-RT. Estimate of the muon passing time in layer 0 falls out of time-consistency window so the hit is not counted in the voting. (d) Overview of the searching grids. Each grid is centered at its former search winner. Grey lines represent the Legendre transform results in the final step.
    }
    \label{fig:algorithm_steps}
\end{figure}

The valid event candidate is processed after the acquisition window closes. For each hit tube, we use the wire-center coordinates $(x_i,z_i)$ in the chamber's cross-sectional plane, measured horizontally and vertically from its bottom-left corner. A line passing through the wire center of tube $i$ satisfies $\rho=x_i\cos\theta+z_i\sin\theta$, where $\theta\in[0^\circ,180^\circ)$ is the angle of the line's normal. Because the line fitted to the wire centers of the hit tubes is not expected to align exactly with the true muon trajectory, relatively wide bins are used at this stage. A Monte Carlo simulation of one million tracks yields a width of \(2.2~\mathrm{mm}\) for the distribution of differences in \(\rho\) and a width of $1.2^\circ$ for the distribution of differences in $\theta$ between the true track and the line fitted to the wire centers. As a result, the Legendre histogram is chosen to have 64 bins in $\theta$ ($2.8125^\circ$ per bin) and 96 bins in $\rho$ (\(4~\mathrm{mm}\) per bin). For each value of $\theta$, every hit votes for the corresponding $\rho$ bin and the two neighboring bins on either side, giving an effective $\rho$ bin width of $\pm 10$ mm. After all hits have voted, the centroid of the highest-scoring cells is calculated and used to define the seed, as shown in Fig.~\ref{fig:algorithm_steps}(a). This seed corresponds to a candidate track in geometric space, as illustrated in Fig.~\ref{fig:algorithm_steps}(b). Only the two hits closest to this candidate track in each layer are retained for subsequent reconstruction.

Each selected hit provides an estimate of the muon passing time, $\tau_i=t_i-t_{\mathrm{RT}}(d_i)$, where $t_i$ is the measured hit arrival time,  $d_i$ is the distance of the tube center to the seed track, and $t_{\mathrm{RT}}$ is the drift time obtained from the inverse $R(t)$ relation. If $ d_i$ is larger than the tube radius, the corresponding hits will be excluded for $t_0$ voting. In our implementation, $t_i$ is further corrected by time-slew correction and channel-specific synchronization offset. If all selected hits originate from the same muon track, their $\tau_i$ estimates should be consistent with a common value of $t_0$ (shown in Fig.~\ref{fig:algorithm_steps}(c)). The candidate $t_0$ is defined as the median of all mutually consistent $\tau_i$ values within a window of $\pm \delta$ ns ($\delta=25$ for the initial seed track). A seed track is considered valid if it receives votes from at least four layers, with contributions from both multi-layers of the chamber. 

The next stage is a grid search in $(\theta, \rho)$ space, as illustrated in Fig.~\ref{fig:algorithm_steps}(d). We first perform a coarse grid search centered on the seed-track parameters obtained from the wire-center positions. The coarse grid contains 7 bins in $\theta$ and 13 bins in $\rho$, spanning $\pm 3^\circ$ and $\pm 6$ mm respectively. At each grid point, the $t_0$-voting procedure is applied using a time-consistency window of $\delta=25$ ns. The best coarse-grid track candidate is determined by maximizing the distinct support layers. If the support layers are the same, we rank them by the multilayer coverage, whether the hits cover both sides of the track, and then a smaller $\tau_i$ spread. 

A fine-grid search is then performed around the best track candidate from the coarse search. This fine grid contains 9 bins in $\theta$ and 15 bins in $\rho$, spanning $\pm 1^\circ$ and $\pm 1.75$ mm, respectively. The time-consistency window is reduced to $\delta=6.25$ ns since the refined track parameters provide more precise estimates of $t_0$. The winning track candidate is determined by the same strategy as the coarse grid search. 

For the winning fine-grid search track candidate, $t_0$ is further refined. In each layer, we select the hit that satisfies $|d_i-R(t_i-t_0)|\leq 1.20$ mm, where $R(t_i-t_0)$ is the drift radius obtained from the $R(t)$ relation. If the selected hits span at least four layers, $t_0$ is updated to the mean of their individual passing-time estimates. The refined $t_0$ is then fixed, and the drift radius of each hit is calculated as $r_i=R(t_i-t_0)$. 

A $5 \times 13$ polishing grid is centered on the track parameters obtained from the fine-grid search. It has 5 $\theta$ points with a step size of 0.125$^\circ$ and 13 $\rho$ points with a step size of $0.125~\mathrm{mm}$. For each candidate on the polishing grid, the residual of hit $i$ is defined as $\Delta_i=d_i-r_i$. In each detector layer, only the hit with the smallest $|\Delta_i|$ is retained. A hit is classified as an inlier if $\lvert\Delta_i\rvert<1.20~\mathrm{mm}$. In the accumulator, each valid hit votes for both tangent branches of its drift circle in Legendre space, and thus every grid point accumulates contributions from curves lying within half a grid spacing, corresponding to $\pm 0.0625~\mathrm{mm}$. For each grid point, we count the number of distinct detector layers contributing votes to that point and also its two adjacent $\rho$ points. The final candidate is selected by maximizing the support layers. Ties are resolved, in order, by the number of layers with inlier hits, multilayer coverage, the residual, and the number of raw votes at the candidate’s own $\rho$ point, with larger values preferred except for the residual, for which a smaller value is preferred. 

The firmware assigns one of the following outcome codes to each event: \texttt{OK}, if a track segment is reconstructed with a valid $t_0$; \texttt{NO\_T0\_CONSENSUS}, if a segment is found but no common $t_0$ can be determined for its associated hits; \texttt{NO\_GLOBAL\_SEED}, if no peak is identified in the $(\theta,\rho)$ parameter space; and \texttt{RESOURCE\_OR\_NUMERIC\_ERROR}, if more than a predefined number of hits are recorded within the allowed time window or the event comes too fast to have available reconstruction core for it. The processing time, or the latency, $\delta t_{\text{lat}}$, is also measured and recorded.

\section{Performance of the self-triggering and segment-fitting algorithm}
To study the performance of the algorithm, cosmic-muon events are recorded in two independent runs: one using the scintillator as an external trigger and the other using the sMDT self-triggering algorithm. In the self-triggered run, the scintillator signal is also recorded to provide an independent reference measurement to determine the cosmic-muon arrival time, but it does not contribute to the trigger decision. 

\subsection{Offline track reconstruction}
The trigger efficiency and the $t_0$ and spatial resolutions are evaluated using offline-reconstructed tracks as the reference. The offline reconstruction proceeds as follows. An event is initially selected if it contains 6 to 12 tube hits, with at least one hit in each multilayer. Offline track candidates are seeded by considering all possible left-right assignments of each drift circle. For each assignment, hit $i$ is represented by a point displaced horizontally by $\pm r_i$ from the corresponding wire center. A straight line is then fitted to these points using the weighted least-squares method. The weights are set to be $1/\sigma^2(r_i)$, where $\sigma(r_i)$ denotes the single-tube resolution at drift radius $r_i$. Each candidate is ranked according to the sum of its squared pulls, $\chi^2=\sum_i [d_i-r_i)/\sigma(r_i)]^2$, where $d_i$ is the distance between the fitted line and the center of wire $i$. Depending on the size of the candidate track pool, the worst 30\%-75\% of
candidates are discarded, while at least 16 candidates are retained. The remaining candidates are refined in two passes. In each pass, the event time offset, $\Delta t_0$, is first corrected using one Gauss-Newton step, after which the drift radii are recomputed. A candidate is updated only if the procedure improves its score. After each pass, the candidates are sorted again, and the worst ones are removed.

The highest-ranked offline track candidate is refined with two additional passes, followed by a final fit to minimize the $\chi^2$ defined above. The fit is iterated until the changes in the fitted angle, intercept, and $\Delta t_0$ are smaller than $10^{-3}$ rad, $10^{-3}$ mm, and $10^{-3}$ ns, respectively, or until the maximum of 50 iterations is reached. During each iteration, a hit is included only if its distance from the fitted line is less than 7.1 mm and its residual satisfies $|d_i-r_i| < 5\sigma(d_i)$, where $\sigma(d_i)$ is the corresponding single-tube resolution evaluated at the track-to-wire distance $d_i$. The reconstructed track is accepted if at least six hits satisfy these criteria in the final fit.

\subsection{Trigger efficiency}

\begin{figure}[pos=ht]
    \centering
    \includegraphics[width=0.8\linewidth]{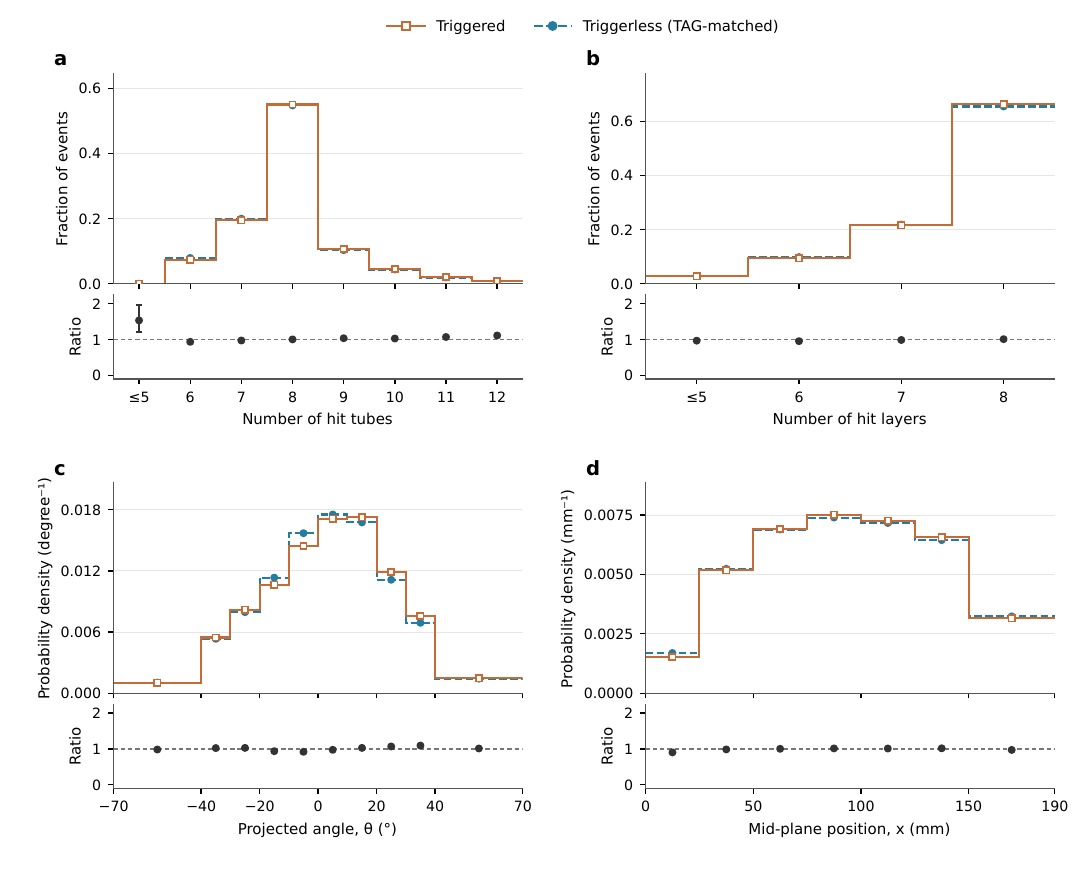}
    \caption{Hit and track distributions comparison between original triggered firmware and Legendre-based triggerless firmware.}
    \label{fig:hit_track_distribution_main}
\end{figure}

The distributions obtained from externally-triggered and self-triggered cosmic-muon runs are compared in Fig.~\ref{fig:hit_track_distribution_main}. The comparison includes the numbers of hit tubes and hit layers, the projected track angle, and the track position at the chamber midplane. Each distribution is normalized to unit area. Overall, the two trigger configurations yield good agreement.

Table~\ref{tab:efficiency_data} presents a detailed breakdown by outcome code of sMDT self-triggered cosmic muon events with both a valid scintillator signal and an offline-reconstructed track candidate. Of these events, $(99.39 \pm 0.02)\%$ are triggered and assigned a valid $t_0$, demonstrating the high trigger efficiency of the algorithm in identifying cosmic muon events.

\begin{table}[pos=ht]
  \centering
  \caption{Breakdown of the outcome code of sMDT self-triggered cosmic muon events containing a valid scintillator signal and an offline-reconstructed track candidate.}
  \label{tab:efficiency_data}
  \begin{tabular}{lr}
    \toprule
    Output & Fraction (\%) \\
    \midrule
    \texttt{OK}   & $99.39 \pm 0.02$ \\
    \texttt{NO\_T0\_CONSENSUS} & $0.58 \pm 0.02$   \\
    \texttt{RESOURCE\_OR\_NUMERIC\_ERROR} & $0.026 \pm 0.004$ \\
    \texttt{NO\_GLOBAL\_SEED}  & $<0.001$                            \\
    \bottomrule
  \end{tabular}
\end{table}

Because the scintillator and sMDT detector have different geometrical acceptances, the absolute trigger efficiency of the sMDT self-triggering algorithm cannot be determined directly from the two data sets. Instead, it is estimated using a Monte Carlo simulation. Truth-level tracks passing through the sMDT chamber are generated, and the corresponding drift radii are calculated. The drift radii are then smeared with a Gaussian resolution of 100 $\mu$m, after which the Legendre-based trigger algorithm is applied. For a sample of 100,000 muon tracks, the absolute trigger efficiency is $99.62 \pm 0.02$ \%. The efficiency exhibits a slight dependence on the number of detector layers traversed by the muon; the results for tracks crossing 6, 7, and 8 layers are shown in Table~\ref{tab:truth_mc_valid_t0}.

\begin{table}[pos=ht]
  \centering
  \caption{Simulated overall absolute trigger efficiency of the Legendre-based trigger algorithm for 100,000 muon tracks. Detailed breakdown of the results for tracks crossing 6, 7, and 8 detector layers are also shown.}
  \label{tab:truth_mc_valid_t0}
  \begin{tabular}{lrc}
    \toprule
    & Tracks & Trigger efficiency (\%) \\
    \midrule
    All tracks & 100,000 & $99.62 \pm 0.02$ \\
    \midrule
    6 layers & 11,552 & $97.59 \pm 0.15$ \\
    7 layers & 22,305 & $99.91 \pm 0.02$ \\
    8 layers & 66,143 & $99.88 \pm 0.02$ \\
    \bottomrule
  \end{tabular}
  \par\smallskip
\end{table}

\subsubsection{$t_0$ and segment-direction resolutions}
To evaluate the resolutions of the muon arrival time, $t_0$, and the segment direction, $\theta$, the values determined by the online sMDT self-triggering algorithms are compared with those obtained from offline reconstruction. The online and offline track parameters are also extrapolated to the chamber midplane, and the resulting $x$-positions are compared. The difference distributions are shown in Fig.~\ref{fig:resolution}(a) for the $t_0$ resolution, Fig.~\ref{fig:resolution}(b) for the $ \theta$ resolution, and Fig.~\ref{fig:resolution}(c) for the midplane position resolution. Each distribution is modeled as the sum of two Gaussian components and the amplitude-weighted width is defined as $\sigma_{\mathrm{amp}}=(A_n\sigma_{\mathrm{narrow}}+A_w\sigma_{\mathrm{wide}})/(A_n+A_w)$, where $A_n$ and $A_w$ are the peak amplitudes of the narrow and wide Gaussian components, respectively. The double Gaussian model is fitted to the central 95\% of each distribution. The average amplitude-weighted width is found to be $1.468 \pm 0.007$ ns for $t_0$, $1.048 \pm 0.004$ mrad for $\theta$, and $66.53 \pm 0.22$ $\mu$m for the track position at the chamber midplane.

\begin{figure}[pos=ht]
\centering
\includegraphics[width=0.9\linewidth]{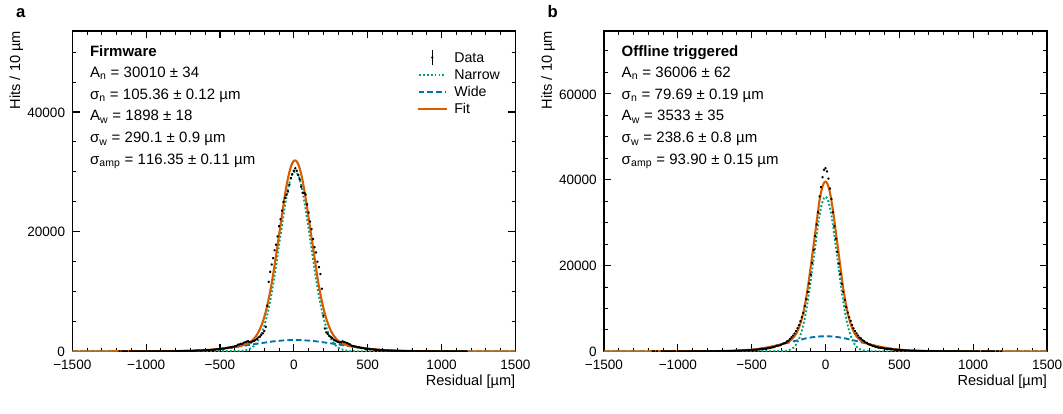}
\caption{Biased fit residual for triggerless firmware fit results and offline triggered fit results.}
\label{fig:fitresidual}
\end{figure}

The single-tube spatial resolution is evaluated using the fit residuals, defined as the measured drift radius minus the distance from the wire center to the reconstructed track. Because the online algorithm performs only a single fit using all associated hits, unbiased residuals cannot be obtained. Biased residuals are therefore used for a consistent comparison. Their distributions are shown in Fig.~\ref{fig:fitresidual}(a) for the algorithm implemented in firmware and Fig.~\ref{fig:fitresidual}(b) for the offline reconstruction. The corresponding amplitude-weighted widths are \(116.35 \pm 0.11~\mu\mathrm{m}\) and \(93.90 \pm 0.15~\mu\mathrm{m}\), respectively. These results show that the real-time track and hit reconstruction achieves a precision close to that of the offline reconstruction.

\begin{figure}[pos=ht]
\centering
\includegraphics[width=0.9\linewidth]{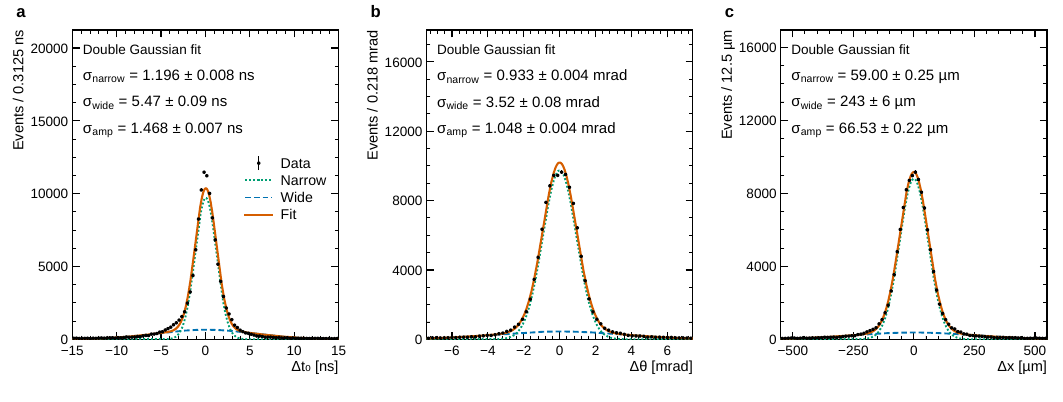}
\caption{$t_0$ and segment-direction resolutions: (a) $t_0$ resolution; (b) Track direction $\varphi$ resolution; and (c) Midplane position $x$ resolution.}
\label{fig:resolution}
\end{figure}

\subsubsection{Latency}
\begin{figure}[pos=ht]
    \centering

    \begin{subfigure}[b]{0.48\linewidth}
        \centering
        \includegraphics[width=\linewidth]{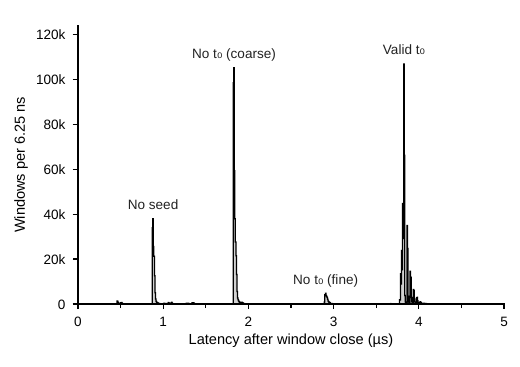}
        \caption{}
        \label{fig:first}
    \end{subfigure}
    \hfill
    \begin{subfigure}[b]{0.48\linewidth}
        \centering
        \includegraphics[width=\linewidth]{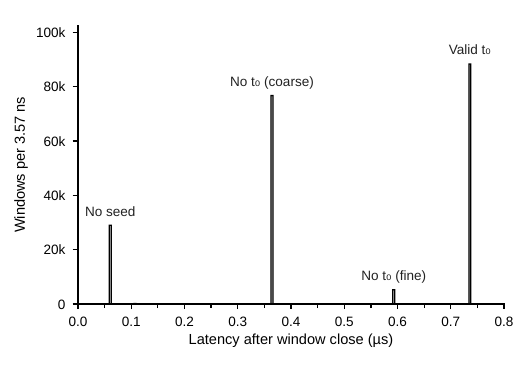}
        \caption{}
        \label{fig:second}
    \end{subfigure}

    \caption{Latency Distribution: (a) Firmware latency tested on current miniDAQ; and (b) Simulated latency on XCKU15P.}
    \label{fig:latency_comparison}
\end{figure}

Figure~\ref{fig:latency_comparison} shows the latency distribution for events processed by the self-triggering algorithm. The latency depends on the stage at which an event exits the pipeline. Events for which no global seed is identified in the initial Legendre parameter space search are rejected early, with a typical latency of 0.88 $\mu$s. Events with a valid seed but no consistent $t_0$ solution in the coarse-grid search require further processing and are rejected after approximately 1.83 $\mu$s. If no consistent $t_0$ solution is found in the fine-grid search, the event is rejected after approximately 2.91 $\mu$s. For events that satisfy all selection and quality requirements, the complete online reconstruction of a high-quality track candidate, including the determination of $t_0$ and segment direction, requires approximately 3.83 $\mu$s, and 99\% of them finish within 4.02 $\mu$s. 

The degree of parallelism is constrained by the resources available on the FPGA used in the miniDAQ board. The small fluctuations between 0.4 and 1.5 $\mu$s correspond to events rejected without reconstruction. In most cases, events have more than the maximum allowed number of hits (24 in this case), causing the reconstruction to terminate and return a numerical-error flag. An event may also arrive when all reconstruction cores are occupied, in which case it will not be processed. In both situations, the event is assigned an outcome code of \texttt{RESOURCE\_OR\_NUMERIC\_ERROR}. The latency can be reduced by omitting the coarse- and/or fine-grid searches at the cost of degraded $t_0$ and spatial resolutions.

The measured hardware latency is reproduced in the RTL simulation. For a fully reconstructed window containing $N$ hits, the simulation predicts a latency of $604+N$ clock cycles from the close of the acquisition window to the availability of the final track parameters. There will be five additional cycles for each hit beyond eight hits used in the final polish step. If the event has 8 hits, it corresponds to a typical latency of 3.83 $\mu$s (612 cycles), consistent with the median latency measured with the miniDAQ FPGA board. 

\subsection{Resource Report}
Table~\ref{tab:resource report}(a) summarizes the resource utilization of the firmware implemented on the miniDAQ board that uses the Xilinx Kintex UltraScale XCKU035 FPGA. The current design uses approximately 85\% of the device's lookup tables and operates at a clock frequency of 160 MHz. 

Because the RTL simulation reproduces the board behavior cycle by cycle, it can also be used to predict the performance of the same algorithm on other FPGA platforms. We therefore implemented the algorithm on the larger Xilinx Kintex UltraScale+ XCKU15P FPGA used on the L0MDT board for the ATLAS HL-LHC MDT upgrade~\cite{Cieri:2020bfv}. Its additional resources enable greater computational parallelism and increase the reconstruction clock frequency to 280 MHz. The resulting design uses nearly twice the resources and reduces the latency by a factor of 5.5. Because an XCKU15P evaluation board is not available for this study, its latency is evaluated using RTL simulation with the events recorded by the current firmware. The event gap times are set to be 1.6 $\mu$s, slightly larger than the window time (600 ns). For events within the hit limit, each outcome takes a fixed time after the window closes, and it is independent of the number of hits. 

\begin{table*}[pos=ht]
  \centering
  \caption{FPGA resource utilization: (a) firmware implemented on the XCKU035 FPGA; and (b) same algorithm with more parallel computation implemented on the XCKU15P FPGA.}
  \label{tab:resource report}
  \small
  \begin{minipage}[t]{0.49\textwidth}
    \centering
    \begin{tabular}{lrrr}
      \multicolumn{4}{c}{(a)} \\
      \toprule
      Resource & Used & Available & Utilization \\
      \midrule
      LUT  & 173062 & 203128 & 85.20 \% \\
      FF   & 148285 & 406256 & 36.50 \% \\
      BRAM & 303.5  & 540    & 56.20 \% \\
      DSP  & 165    & 1700   & 9.71 \% \\
      \bottomrule
    \end{tabular}
  \end{minipage}\hfill
  \begin{minipage}[t]{0.49\textwidth}
    \centering
    \begin{tabular}{lrrr}
      \multicolumn{4}{c}{(b)} \\
      \toprule
      Resource & Used & Available & Utilization \\
      \midrule
      LUT  & 370227 & 522720  & 70.83 \% \\
      FF   & 275214 & 1045440 & 26.33 \% \\
      BRAM & 230    & 984     & 23.37 \% \\
      DSP  & 329    & 1968    & 16.72 \% \\
      \bottomrule
    \end{tabular}
  \end{minipage}
\end{table*}

\section{Conclusions}
We present a fast self-triggering and segment-fitting algorithm that identifies cosmic muons and reconstructs their directions using an sMDT chamber without an external trigger detector. The algorithm first uses the wire centers of all hit tubes to obtain an initial estimate of the track parameters. It then performs coarse- and fine-grid searches around this estimate. For each track candidate, the algorithm calculates the distances from the candidate track to the wire centers and converts them into drift times using the inverse $R(t)$ relation. The drift time of each hit provides an estimate of the muon arrival time, $t_0$. A $t_0$-voting procedure using all associated hits identifies the most likely track candidate. The algorithm is implemented in FPGA firmware on the current ATLAS MDT miniDAQ board and is tested with cosmic muons using an sMDT chamber in our laboratory. Its efficiency and resolution are measured and compared with those obtained from offline reconstruction. The latency is also measured both on the hardware and through simulation. Our results demonstrate that a standalone sMDT chamber can trigger on cosmic muons and determine their directions. Our algorithm provides a solution for chamber commissioning without an external trigger detector and potentially MDT-based triggers for charged massive stable particle searches at the HL-LHC.

\section*{Acknowledgement}
The work is supported by the US Department of Energy (DOE) under contract DE-SC007859 (DOE).

\bibliographystyle{model1-num-names}

\bibliography{refs}

\end{document}